# Enhancement of the magnetorheological effect using discontinuous shear thickening with a mixture of magnetic and non magnetic particles

Bossis Georges
*UMR7010, Inphyni,*
*University Côte d'Azur, CNRS*
*17 rue Julien Lauprêtre,06200 Nice, France*
Georges.BOSSIS@unice.fr

Grasselli Yan
*UMR7010, Inphyni,*
*University Côte d'Azur, SKEMA Business School*
*60 rue Dostoievski – CS30085*
*06902 Sophia Antipolis, France*
yan.grasselli@skema.edu

Volkova Olga
*UMR7010, Inphyni,*
*University Côte d'Azur,CNRS*
*17 rue Julien Lauprêtre,06200 Nice, France*
Olga.VOLKOVA@unice.fr

***Abstract*— The phenomenon of discontinuous shear thickening (DST) is observed in suspensions of solid particles with a very high-volume fraction. Using a suspension of magnetic particles at high-volume fraction we have, for the first time, demonstrated experimentally the presence of a frictional network of particles by a simultaneous measurement of the electrical resistance of the suspension and of the rheological curve (Ashtiani et al., 2015). An obstacle for the applications of MRS remains the sedimentation and aggregation linked to the high density of the iron microparticles. We have recently found that a small fraction of ferromagnetic particles within a concentrated suspension of calcium carbonate particles is not only sufficient to trigger the transition by application of a magnetic field but also optimum for the amplitude of the jump in viscosity observed in DST. We shall explain this result from stress transmission between the magnetic particles and nonmagnetic ones, both inserted in the percolated network of particles. We will also highlight the role of the polyelectrolyte adsorbed on the surface of the particles in the DST transition through the balance between the repulsive entropic force between two layers of polymer adsorbed on different particles and the magnetic attraction force.**

***Keywords*— *Magnetorheology, suspension, Rheology Discontinuous shear thickening,.***

## I. INTRODUCTION

Magnetorheological suspensions (MRS) belong to the class of smart materials since the application of a magnetic field can induce a change of phase from the liquid state to the solid state. This solid state is mainly characterized by a yield stress, $\tau_y$ which represents the stress needed to break the solid network of particles created by the application of the magnetic field. When the MRS begins to flow, it is then characterized by a plastic viscosity $\eta_{pl}$ and the simplest rheological law is the Bingham one: $\tau = \tau_y + \eta_{pl}\dot{\gamma}$ where $\dot{\gamma}$ is the shear rate. A commercial MRS with a volume fraction of carbonyl iron particles between 30% and 40% has a typical maximum yield stress of 40-60kPa for a magnetic field of about 400 kA/m corresponding to an induction field of 0.5 Tesla. The two main drawbacks of MRS are the irreversible aggregation of the suspension over time and the weight and size of the magnetic circuit necessary to produce a high magnetic field. The aggregation of the microparticles can be due both to their sedimentation and to the oxidation and erosion of their surfaces. Usually, the surface of the particles is protected from coming in close contact by a coating layer giving a short-range repulsive force between the particles. Depending on the nature of the suspending fluid, it can be a ionic or a polymer layer or both in the case of polyelectrolytes. Different ways to reduce sedimentation and aggregation, keeping a high enough yield stress, are described in some reviews (Ashtiani et al., 2015; Morillas and de Vicente, 2020 , Chauhan et al., 2024; Choi et al., 2024). Another way is to use bidisperse suspensions, or more precisely a mixture of nanoparticles and microparticles; cf (Choi, 2022; Nejatpour et al., 2024). For instance, due to their permanent magnetisation, magnetite nanoparticles are attracted towards the surface of soft carbonyl iron (CI) microparticles but, because of their nanometric size, remain free to move in a shell surrounding the microparticles, thus preventing their aggregation and to some extent, their sedimentation (Burguera et al., 2008; Chaudhuri et al., 2005; López-López et al., 2006; Wereley et al., 2006). So far, it was thought that both types of particles needed to be magnetic since non-magnetic particles would not contribute to the field induced yield stress. In practice this is not necessarily true; for instance it was shown experimentally that adding hollow glass spheres to a suspension of CI particles can increase the field induced yield stress (Ulicny et al., 2010). This effect comes from the combination of the increase of the total volume fraction and of the formation of elongated aggregates under the magnetic attractive forces between CI particles. It was demonstrated by numerical simulation, showing the network of compressive forces extending throughout the cell and including both magnetic and nonmagnetic particles (Wilson and Klingenberg, 2017). This effect opens the way to the use of appropriated mixtures of magnetic and nonmagnetic particles which can be more easily stabilized by the same fluidizer molecules and helps to reduce their average density. Besides, increasing the average volume fraction, lead first to a continuous shear thickening (CST) effect characterized by a viscosity which increases with the shear rate. At still a higher volume fraction, a phenomenon of jamming, called discontinuous shear thickening (DST); can appear and be equally used to improve the MR effect. The DST effect was discovered a long time ago with suspensions of quartz particles in water (Freundlich and Röder, 1938). It manifests by a sudden decrease of the shear rate during a ramp of stress due to the onset of frictional contacts between the particles upon the action of the shear stress which overcomes the repulsive force between them as demonstrated by numerical simulations; a drawing of the evolution of the frictional contacts with the increase of the shear rate can be found in (Mari et al., 2014; Seto et al., 2013, 2013). The existence of this jamming transition in MR suspension of CI particles was observed only in the presence of a high magnetic field even at the low volume fraction of 25% and the critical shear rate was increasing with the magnetic field, although remaining lower than 1 $s^{-1}$ (Tian et al., 2010). For a bidisperse suspension made of nanometric fumed silica and CI particles, Zhang et al (Zhang et al., 2008) have obtained a DST transition at a high critical shear rate which was slightly decreasing with the magnetic field and unexpectedly, was stronger for the smallest proportion of iron particles; we shall see the explanation in a next section. In these shear thickening suspensions the primary effect comes from the shear thickening of the base fluid often made of silica nanoparticles in polyethylene glycol. The formation of linear aggregates of CI particles under the application of the magnetic field can split the silica aggregates resulting in the decrease or even cancellation of the shear thickening effect (Guo et al., 2023). Several other scholars have observed this behaviour in similar shear thickening MR fluids (Li et al., 2014; Qian et al., 2022; Sokolovski et al., 2020; Yan et al., 2024; Zhao et al., 2023). In these experiments no additive molecule was added to coat the particles. Our case is quite different since the base fluid is Newtonian at any shear rate and the discontinuous shear thickening comes from the balance between the repulsive force induced by a coating polymer on the surface of the particles and the compressive forces either hydrodynamic or magnetic. As we shall see in the following, the role of the fluidizer molecule is essential both for the understanding of the rheological behavior and for the control of the viscosity jump versus the shear rate and the magnetic field. In our previous works, we have shown that with a superplasticizer molecule used in cement industry we can obtain huge jump of viscosity in MR suspensions with low magnetic field (10 kA/m<H<30 kA/m) (Bossis et al., 2019, 2016).

After the description of the preparation of the suspension and of the experimental method in section II, we shall see in section III how a joint measurement of the electrical resistance and of the rheological curve can prove the existence of a percolated network of frictional contacts during the jamming transition. The section IV will be devoted to the study of the jamming transition in a bidisperse suspension made of CI particles and of calcium carbonate (CC) particles. The section V will be a discussion of the results based on the role of the superplasticizer molecule and a conclusion.

## II. Materials and methods

In our work, we have used two kinds of particles. The CI particles from BASF (grade HQ) are sold by Imhoff Gmbh have a density measured by gas pycnometer $\rho$=7.7g/cm$^3$. The calcium carbonate (CC) particles from Omya(grade BL200) have a density $\rho$=2.72 g/cm3. A picture of a mixture of these particles is presented in Fig.1. Their size distribution was determined with the help of the software ImageJ on 2300 CI particles and 600 CC particles. Their shape was fitted by an ellipse and their diameter obtained from the average of the major and minor axis. The average radii and standard deviation were $R_{CI}$=0.297µm; $\sigma_{stdCI}$= 0.15µm and $R_{CC}$=1.75µm. $\sigma_{stdCC}$= 0.809µm. Their size distribution is well approximated by a lognormal distribution:

$$P(R) = \frac{1}{R\sigma\sqrt{(2\pi)}} \exp\left[-\frac{(\ln R - \mu)^2}{2\sigma^2}\right]$$

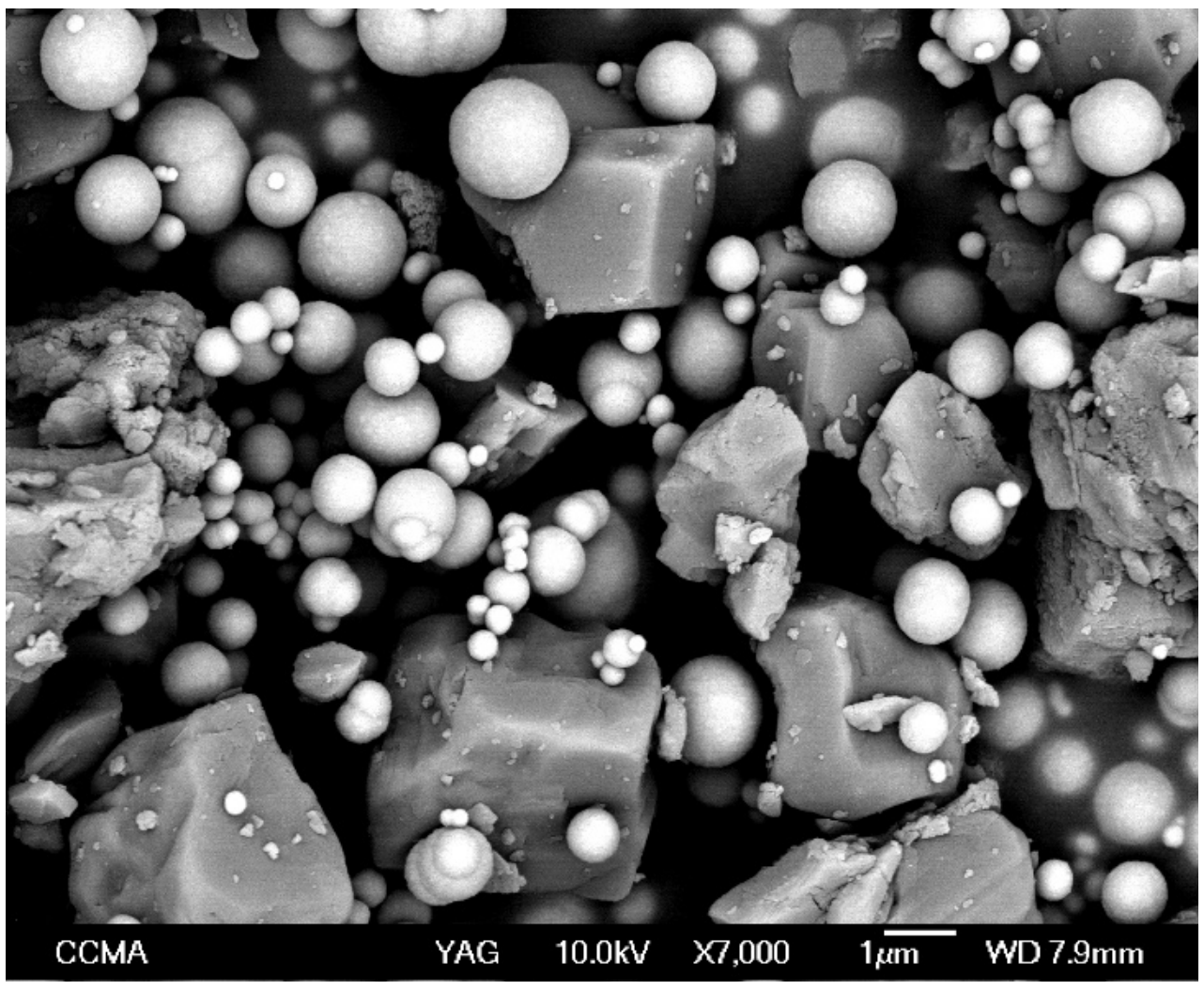

Fig.1 SEM image of CI particles (spherical ones) and CC particles (irregular big ones)

with the following parameters: $\mu$= - 1.292, $\sigma$=0.575 for CI particles and $\mu$= 0.475, $\sigma$=0.4027 for CC particles. The suspending fluid is a mixture of polyethylene glycol (85%) and water (15%) corresponding to a minimum of evaporation. The superplasticizer is a diphosphonate-polyethylene oxide (PEO) whose commercial name is Optima 100; its average length is 44 O-$CH_2CH_2$ groups. The negatively charged phosphonate groups have a sodium counter ion which is replaced by the positively charged iron or calcium atoms when the phosphonate head adsorbs on the particles. The PEO chain acts like a brush polymer which extends in the suspending fluid generating some repulsive force resisting to the interpenetration of two opposite layers when two particles are pushed against each other by the applied shear stress (Bossis et al., 2017). Except if it is specified, the mass of the fluidizer molecule was 0.2% of the total mass of the particles. The rheological measurements were realized on a MCR501 rheometer in plate-plate geometry with a gap around 1mm. These plates were positioned in the middle of a homemade coil which was cooled with water circulation around. The temperature of the sample was maintained at 20°C thanks to a Peltier thermostat below the bottom plate. At this temperature the viscosity of the suspending fluid is $\eta_0$=0.012 Pa.s.

## III. DST with pure CI suspension

Numerical simulations were made considering lubrication forces and contact forces represented by the stiffness of radial and tangential springs when the separation of the surfaces of the particles becomes smaller than a certain value _for instance $10^{-3}$ radius. This way, the authors (Seto et al., 2013) were able to recover the jump of viscosity corresponding to the DST transition and to relate this jump to the formation of a percolating network of frictional contacts. Nevertheless, an experimental proof is not so easy to obtain. Thanks to the use of conductive particles like CI it is possible to check if the DST transition observed on the rheological curve is associated with a jump of resistance of the suspension. We have realized this experiment thanks to a logical output of the rheometer MCR501 allowing to trigger the measurement of the electrical resistance at the same time as the record of the stress and shear rate. For the measurement of the electrical resistance, a bundle of thin copper wires was slightly rubbing on the rotating axis of the upper plate. In Fig. (2) we have plotted the rheograms of a pure CI suspension at a volume fraction $\Phi_{CI}$=0.64 without field and with a small field: H=7.2 kA/m and in Fig.3 the change of resistance associated to the rheogram at H =0 kA/m.

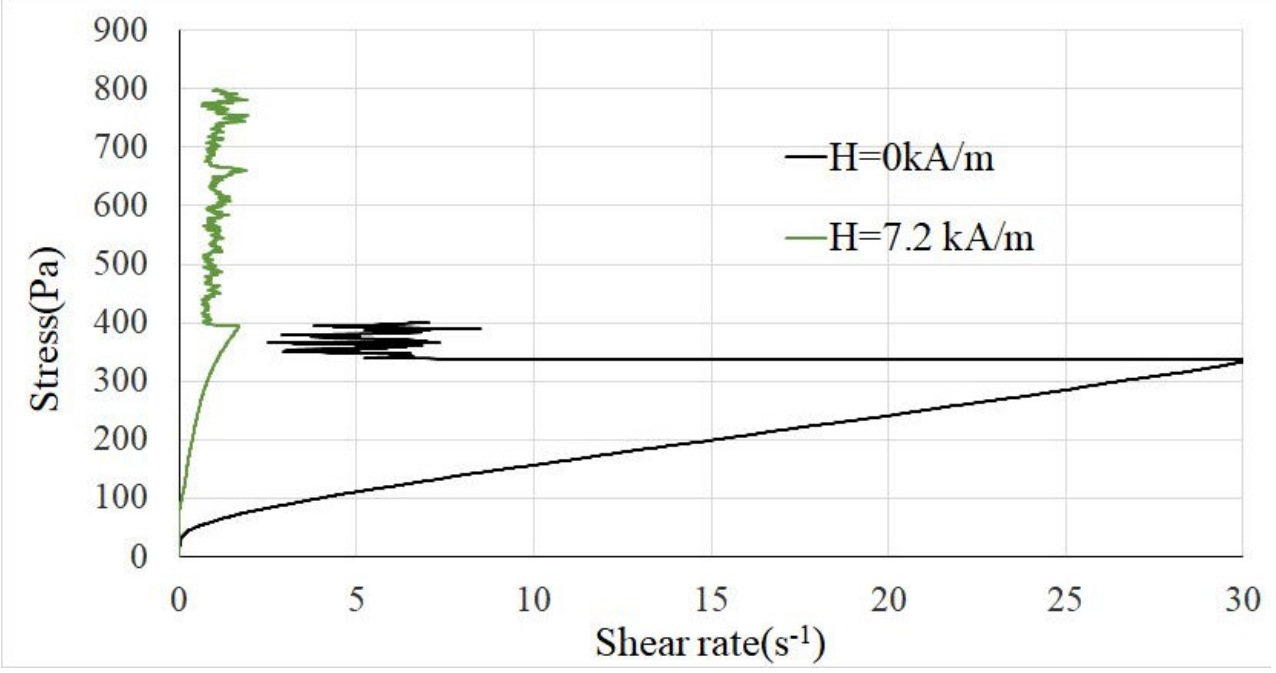

Fig.2 Rheograms for a suspension of pure CI particles at a volume fraction Φ=0.64 at H=0 and H=7.2 kA/m.

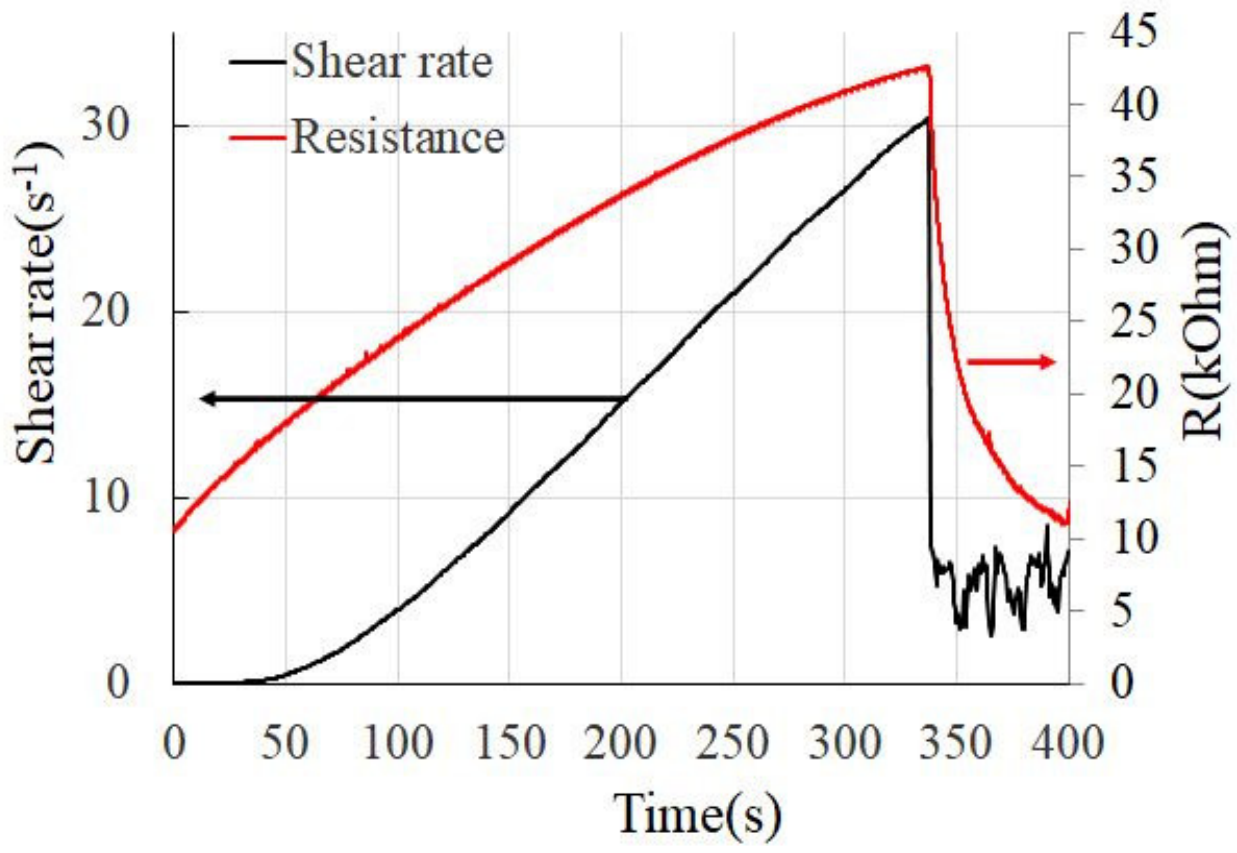

Fig.3 Change of resistance (in red) and of shear rate (in black) during a ramp of stress corresponding to H=0 kA/m.

The drop of resistance (Fig.3) is very well associated to the sudden drop of shear rate which is characteristic of the DST transition during a ramp of shear stress (Fig.2). This is the proof that the transition is due to the formation of a network of percolating contacts between the particles. Note that the resistance is increasing at low shear rate; it means that at rest, after a preshear, some percolating network of particles was already present which is progressively broken by the shear flow until it builds again during the transition. It is also worth noting that, above the critical stress, the suspension continues to flow but at a constant lower shear rate. We have interpreted this behaviour by a relaxation process related to the ejection of the coating layer at a critical stress followed by its return when the flow stops. This back and forth motion of the fluidizer molecule on the surface of the particles results in a motion at a constant shear rate whose value is dictated by the return time of the fluidizer molecule (Bossis et al., 2022). A similar behaviour of the resistance during the ramp of stress at H=7.2 kA/m was obtained. The quantitative interpretation of the DST behaviour was given in (Wyart and Cates, 2014) and subsequent works and some modifications of this model recently proposed in (Bossis et al., 2022). In the following we shall present experimental results obtained on bidisperse suspensions.

## IV. BIDISPERSE SUSPENSIONS

We have done a systematic study of the DST transition of the bidisperse suspension described in section II for different ratios of CI versus CC particles. The ratio in volume of CI particles relative to the total solid content is $F_{CI}=V_{CI}/(V_{CI}+V_{CC})$. The total volume fraction, Φ, is still the volume of solid divided by the total volume. In Fig(4) we have presented the rheograms for $F_{CI}$=5% and Φ=70.5%. First, we see that we have a strong DST transition in the absence of field (blue line) which is not surprising at such high total volume fraction. Second, what is more surprising is the fact that, despite the low ratio in volume of iron particles; $F_{CI}$=5%, we keep a strong effect of the magnetic field even with quite low values of this field. We have fitted the rheograms with two laws: a Bingham law :$\tau = \tau_{yB} + \eta_B \dot{\gamma}$ at low values of the shear rates and a Herschel Buckley (HB) law:$\tau = \tau_{yHB} + K\dot{\gamma}^p$ from a shear rate $\dot{\gamma}_{tr}$ to the critical shear rate of the DST transition: $\dot{\gamma}_c$ . In Table 1 we have reported the values corresponding to the fits of Fig.(4).

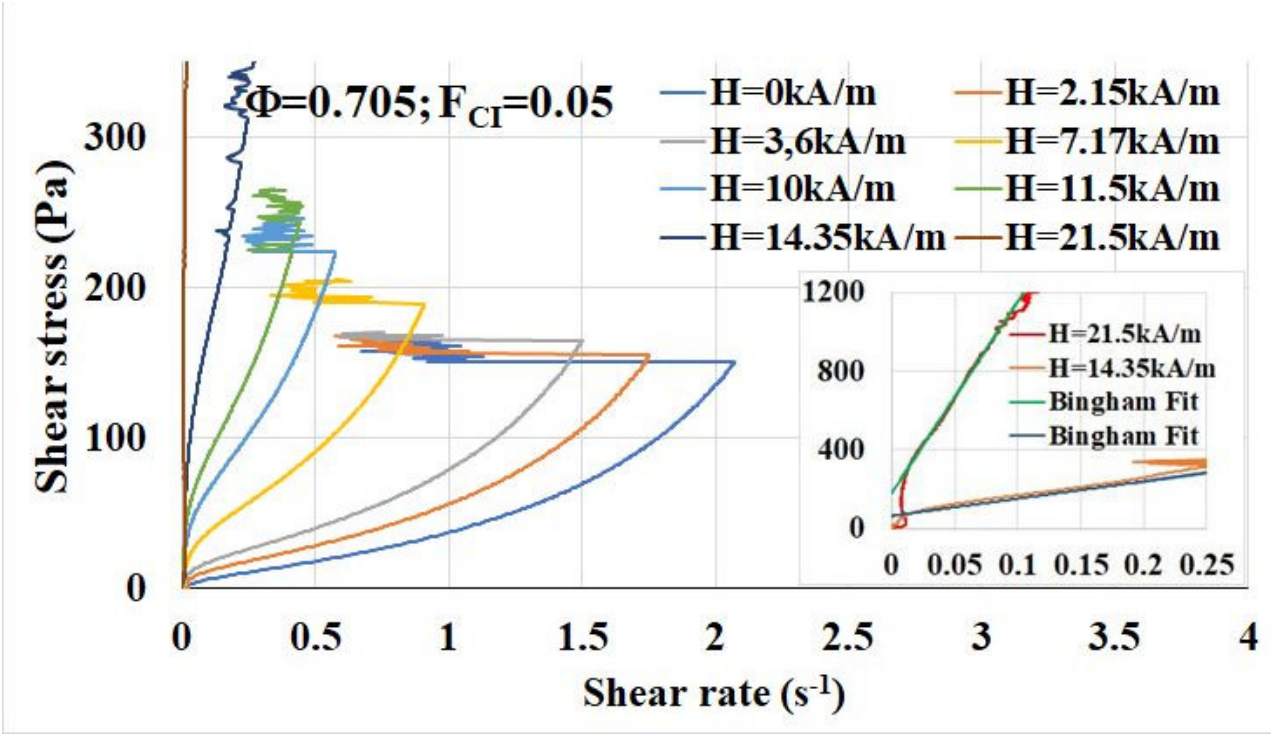

Fig.4 Rheograms for different magnetic fields at $F_{CI}$=0.05 and a total volume fraction Φ=0.705. The insert is a magnification for the two higher fields and the straight lines a fit by the Bingham law.

TABLE I $F_{CI}$=0.05

| H(kA/m) | $\tau_{yB}$ | $\eta_B$ | $\tau_{yHB}$ | K | p | $\dot{\gamma}_{tr}$ |
|---|---|---|---|---|---|---|
| 0 | 1.95 | 31 | 21.6 | 16.1 | 3.42 | 0.8 |
| 2.16 | 5.2 | 44 | 28.7 | 27.8 | 3.38 | 0.7 |
| 3.6 | 8.99 | 56.8 | 34.3 | 49.5 | 3.25 | 0.63 |
| 7.2 | 19.5 | 125 | 48.6 | 222 | 2.6 | 0.4 |
| 10.08 | 29.8 | 252 | 64.2 | 650 | 2.3 | 0.32 |
| 11.52 | 39.14 | 354 | 69 | 880 | 1.91 | 0.2 |
| 12.96 | 50.8 | 524 | 75.2 | 1230 | 1.65 | 0.15 |
| 14.4 | 65 | 877 | 93.2 | 1890 | 1.55 | 0.12 |
| 21.6 | 184 | 9730 | | | | |

In Fig.(5) and Table II, we have the same data still for the total volume fraction Φ=0.705, but for a larger value of $F_{CI}$: $F_{CI}$=0.29. All the values are in MKSA units. We see that for $F_{CI}$=0.29, although the total volume fraction is the same, we do not observe a DST transition even for the higher field. In the penultimate column we have the exponent, p, of the HB law which decreases with the field and even becomes smaller than one (shear thinning behaviour) for $F_{CI}$=0.29.

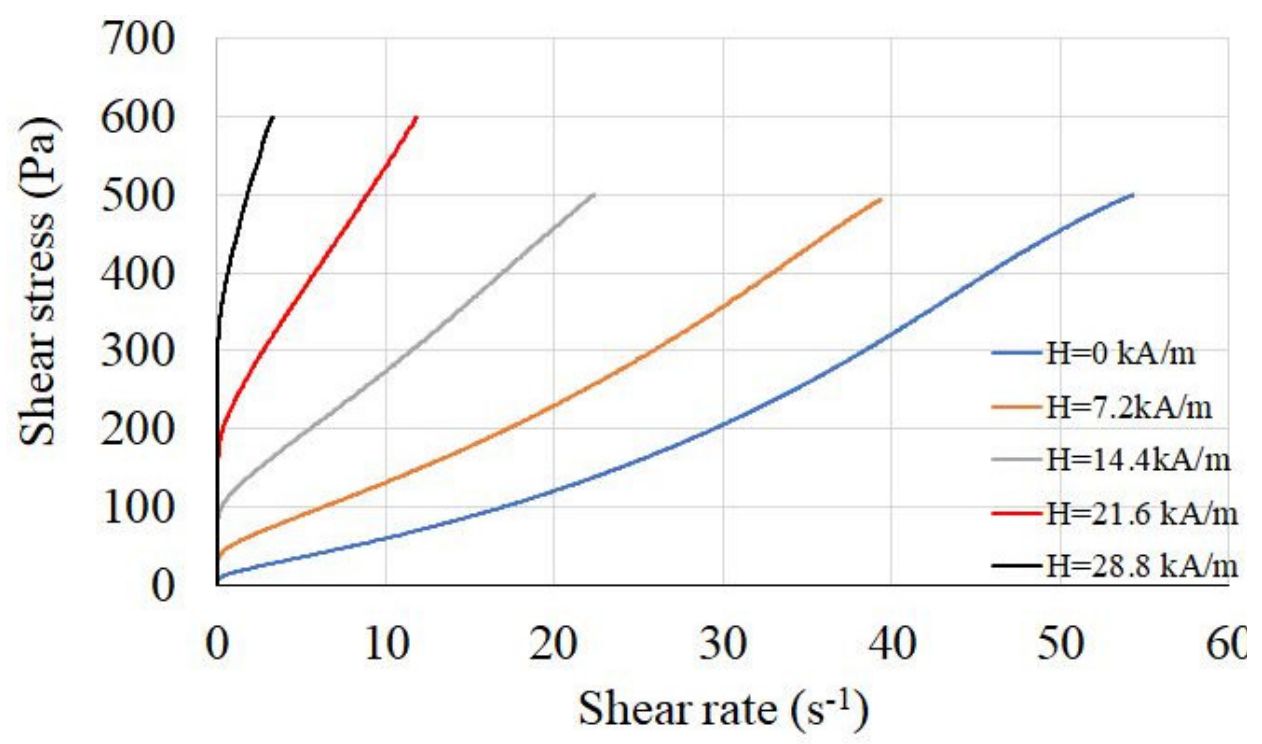

Fig.5 $F_{CI}$=0.29 total volume fraction Φ=0.705. Shear stress versus shear rate for different magnetic fields

TABLE II $F_{CI}$=0.29

| H | $\tau_{yB}$ | $\eta_B$ | $\tau_{yHB}$ | K | p | $\dot{\gamma}_{tr}$ |
|---|---|---|---|---|---|---|
| 0 | 6.6 | 5.2 | 34.7 | 0.32 | 1.88 | 12 |
| 7.2 | 33.9 | 8.69 | 53.5 | 2.3 | 1.44 | 12 |
| 14.4 | 81 | 16.6 | 83.4 | 14.3 | 1.07 | 8 |
| 21.6 | 145.9 | 39.9 | 139 | 44.8 | 0.88 | 1.5 |
| 28.8 | 232 | 149.8 | 217 | 131.6 | 0.63 | 0.2 |

The comparison of the Bingham viscosity, $\eta_B$, versus the field for the two values of $F_{CI}$ is represented in Fig. (6). It is noticeable that the viscosity for $F_{CI}$=0.05 is much larger than for $F_{CI}$=0.29 but above all, that for $F_{CI}$=0.05 the increase of the viscosity with the field is almost three orders of magnitude for a modest field of about 20 kA/m.

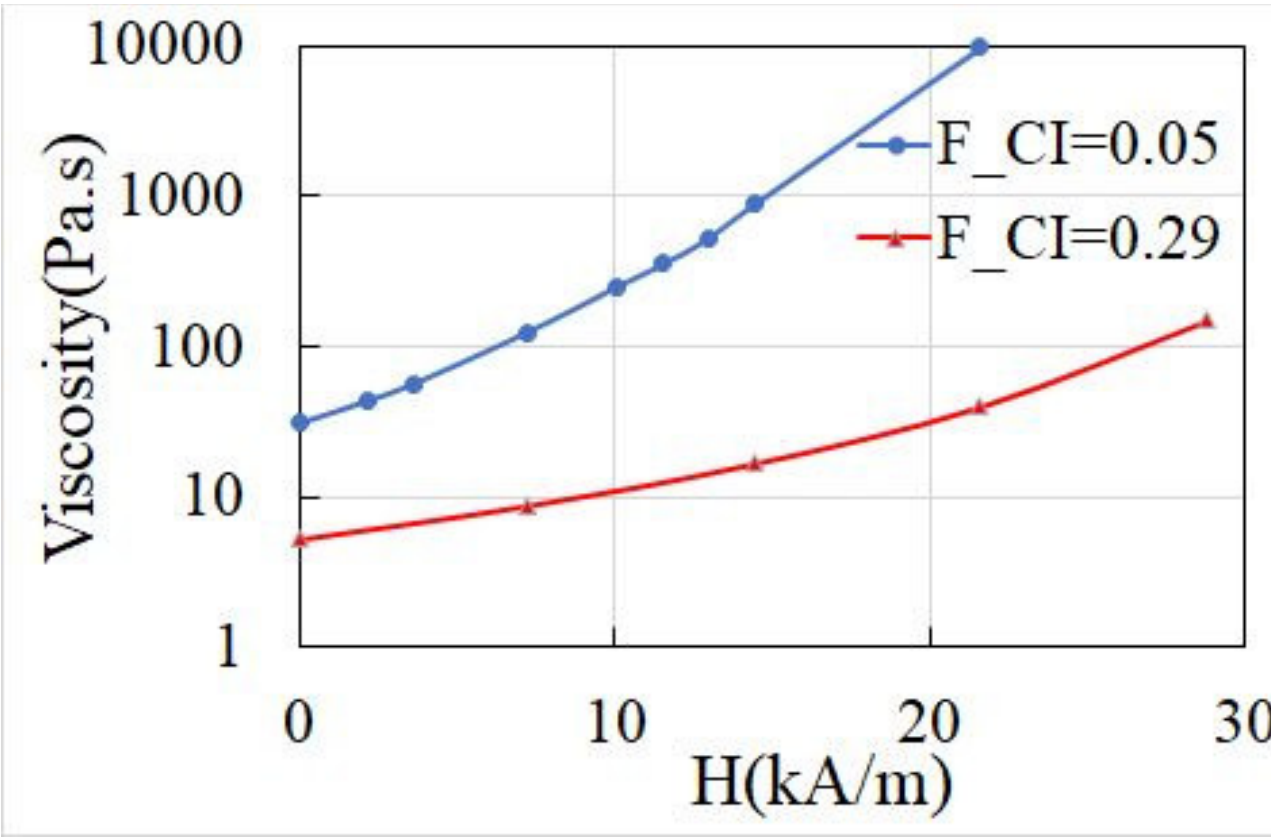

Fig. 6 Bingham viscosity; $\eta_B$, for two ratios of iron particles: $F_{CI}$=0.05 and $F_{CI}$=0.29 and the same total volume fraction; Φ=0.705.

On the same range of field, the increase of viscosity is only a factor of 8 for $F_{CI}$=0.29. The synthesis of the change of the Bingham viscosity with $F_{CI}$ for three different fields is plotted in Fig. (7). What we see first is that the viscosity is smaller for intermediate values of $F_{CI}$. This is a well-known effect in bidisperse suspension with a large ratio of diameters such that the small particles can fill the interstices between the large particles and thus increase the maximum volume fraction, $\Phi_{max}$ corresponding to an infinite viscosity (Pednekar et al., 2018). As the viscosity of concentrated suspension is well represented by the law $\eta \propto (1-\frac{\Phi}{\Phi_{\max}})^{-2}$, it explains the decrease of viscosity for intermediate values of $F_{CI}$.

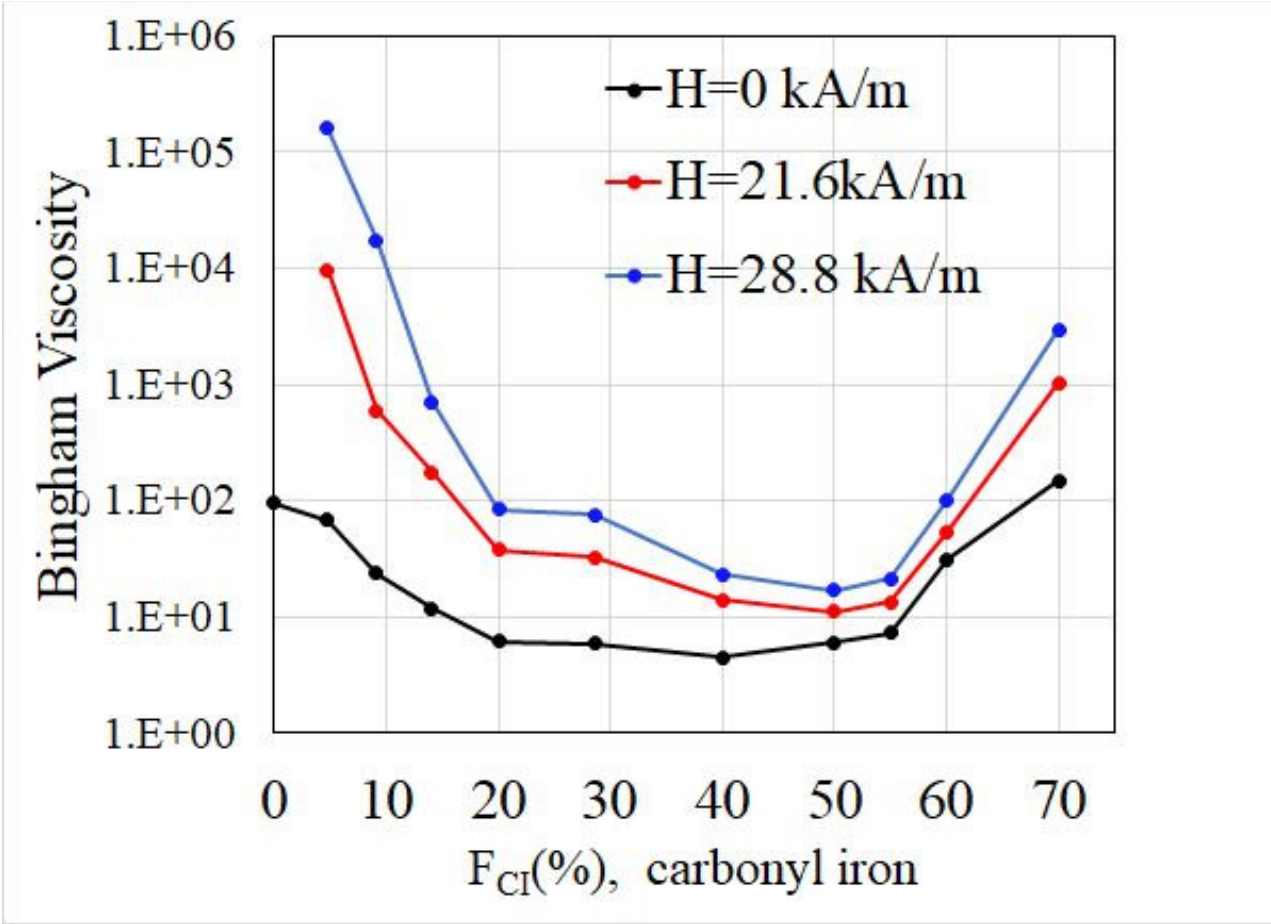

Fig.7 Bingham viscosity versus the percentage in volume of iron particles: $F_{CI}$ for three different fields.

The most intriguing part revealed in Fig. (7) is that the increase of the Bingham viscosity with the magnetic field is much larger at a small volume fraction of iron particles that at larger ones. One would expect that the more magnetic particles, the greater the effect of the field on the viscosity would be. The main explanation for this contradiction comes from the effect of the demagnetizing field in the configuration of a thin sample between two plates since in this case, the internal magnetic field is divided by the magnetic permeability, μ, of the suspension: $H_i=H/\mu$. As the magnetic permeability increases proportionally to the volume fraction, $\Phi_{CI}$ of iron particles: $\mu=1+c.\Phi_{CI}$ with c~8.5 (de Vicente et al., 2002), it gives an important decrease of the effective field acting on the suspension when $F_{CI}$ increases since we have $\Phi_{CI}=\Phi.F_{CI}$. The result is that the internal field is divided by 2.6 between $F_{CI}$=0.05 and $F_{CI}$=0.4 and by 4 between $F_{CI}$=0.05 and $F_{CI}$=0.7. It can explain the decrease of the viscosity by one order of magnitude between $F_{CI}$=0.05 and $F_{CI}$=0.7 if we expect for the viscosity the same behaviour as for the yield stress with a dependence $\propto H_i^{3/2}$ (Ginder et al., 1996), in the range of intermediate field. Nevertheless for the decrease between $F_{CI}$=0.05 and $F_{CI}$=0.29, the role of the decrease of the viscosity with the increase of $\Phi_{max}$ is also important for intermediate values of $F_{CI}$ (Bossis et al., 2023).

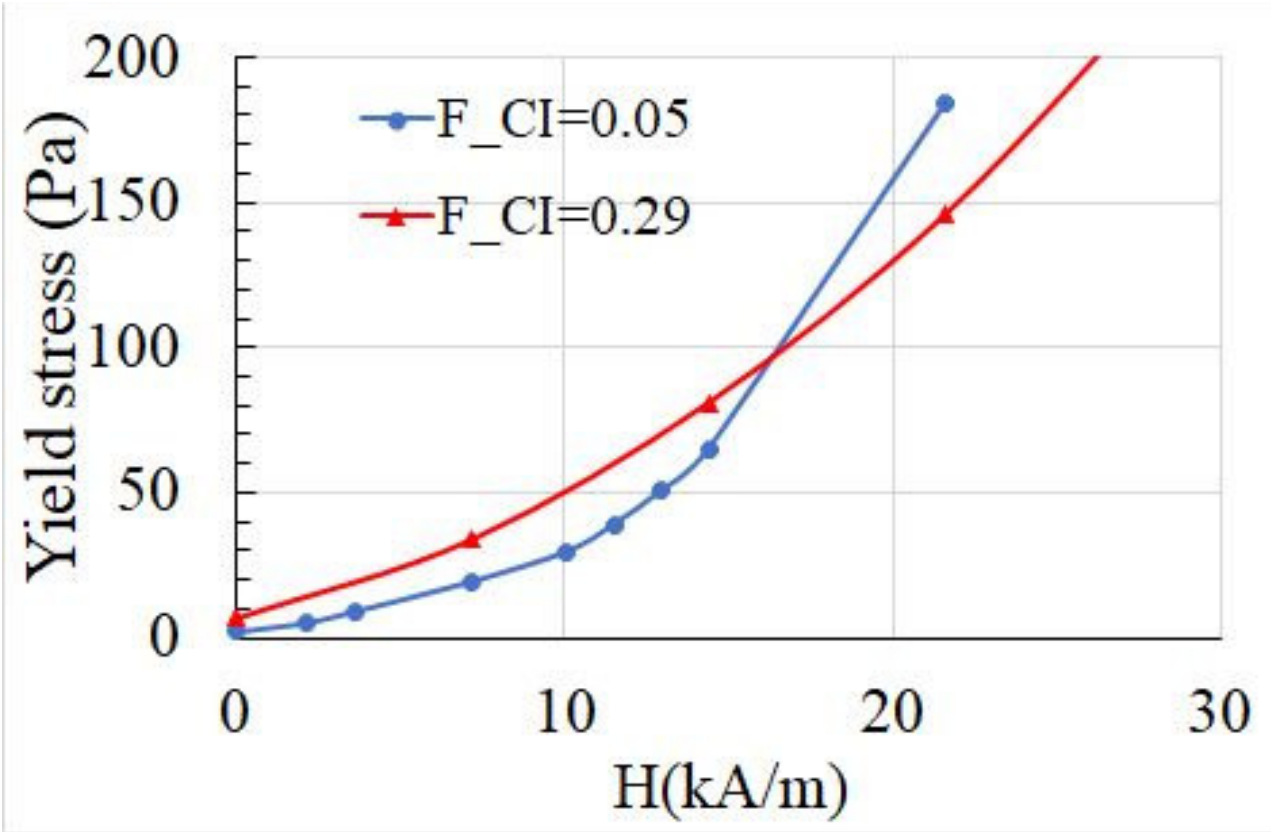

Fig.8 Yield stress, $\tau_{yB}$ for $F_{CI}$=0.05 and $F_{CI}$=0.29

It is worth emphasizing that, contrary to the viscosity, the change of yield stress with the field is quite similar between $F_{CI}$=0.05 and $F_{CI}$=0.29 as we can see in Fig. 8 and we can even see that the yield stress is smaller for $F_{CI}$=0.05 than for $F_{CI}$=0.29 at small field. Indeed, at zero magnetic field, the yield stress is governed by the Van der Waals attraction forces which are inversely proportional to the diameter of the particles; this explains why at zero or small external field it is higher for the greater proportion of small iron particles. This difference of behaviour can also be tentatively explained by the role of the of the coating layer on the particles. To show the influence of the coating polymer on the DST phenomenon we come back to the pure iron particle suspension at zero field. The evolution of the DST with the mass concentration of the fluidizer relatively to the mass of particles is shown in Fig. (9).

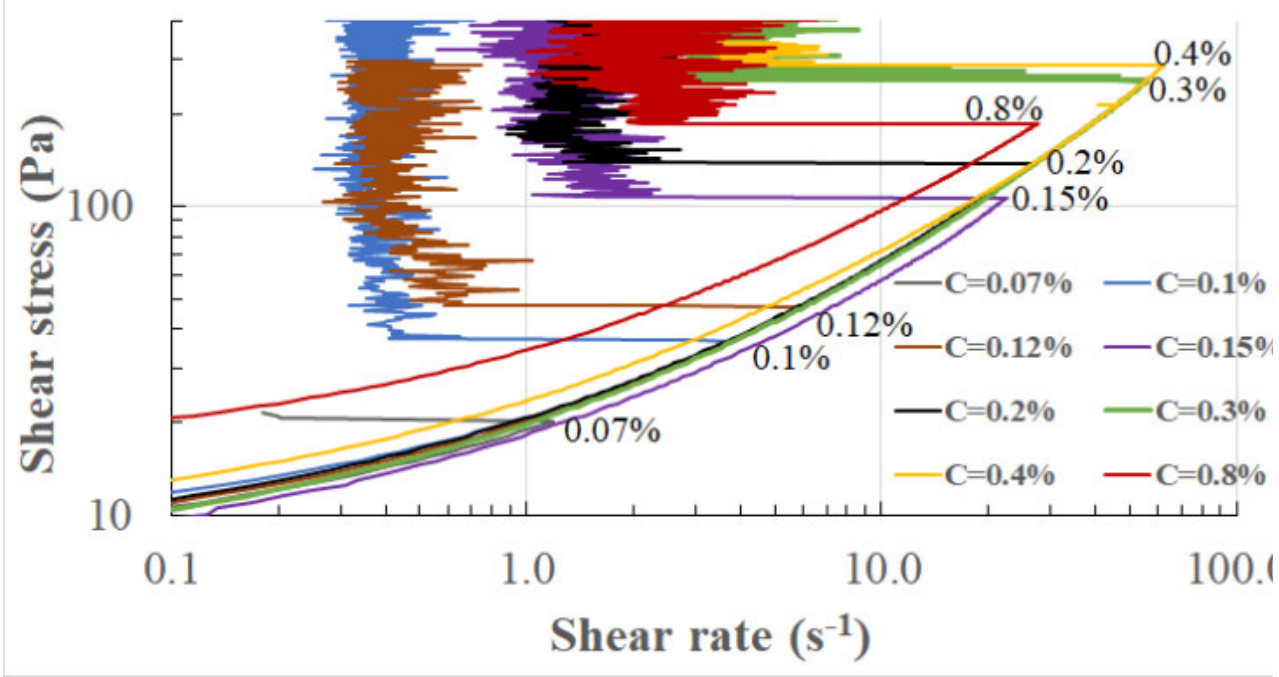

Fig.9 Evolution of the DST with the mass percentage of the plasticizer Optima 100 for a suspension of pure iron at $\Phi$=0.64 and zero field.

For a better lecture, the percentage of fluidizer is indicated next to or above the return of the shear rate during the DST transition. We see that in the range of concentration above 0.07%, the main effect of the increase of the concentration of the fluidizer is to shift the critical shear rate and critical stress at higher values between 0.07% and 0.4%, for the last values we have a saturation of the effect together with an increase of the viscosity. Below 0.07% we did not report the values because we have a strong increase of the yield stress, and the suspension is no longer liquid. It is obvious that the onset of the DST needs a more and more important hydrodynamic stress to happen which is related to the increase of the repulsive force coming from the coating layer of polymer. The critical stress can be evaluated and should increase as $(\delta/s)^3$ (Bossis et al., 2017), where $\delta$ is the thickness of the layer of polymer and s the average distance between two polymers in this layer. We can therefore assume that the completion of the first polymer layer occurs for a mass fraction of 0.4% of polymer and above, the critical stress saturates or even decreases as the thickness of the layer increases. Actually the decrease observed at 0.8% is close to the range of uncertainty but would be compatible with the theoretical prediction of the repulsive force versus the thickness of the coating layer (Bossis et al., 2017). The experiments were made with a concentration of 0.2% in mass corresponding to a layer of Optima 100 which is close to be completed (cf Fig.9)

## V. Discussion

Magnetorheological suspensions made of carbonyl iron particles at intermediate volume fractions (~ 40%) have the advantage to present an off field viscosity around 1 Pa.s at 100 $s^{-1}$ but they need a high induction (B≥0.5 Tesla) to obtain a yield stress around 50kPa. With high volume fractions ($\Phi$>60%), thanks to a superplasticizer used in cement industry, the viscosity remains not too high (2<$\eta$<10 Pa.s for 60%≤$\Phi$≤64% ) and, due to the presence of the DST transition, the jump of stress can attain large values with small induction like 150kPa for B= 0.025T (Bossis et al., 2016). This DST transition is due to the percolation of frictional contacts as can be proved by a simultaneous measurement of the electric resistance and the shear rate during a ramp of stress. We can use still higher volume solid volume fraction ($\Phi$>70%) by mixing two different sizes of particles such that the small particles can be inserted in the gap of a dense network of the bigger particles. Furthermore, the lager particles do not need to be magnetic because, at such a large-volume fraction, field induced aggregation on a small portion of the particles is sufficient to trigger the DST transition. It is interesting to work with a small quantity of magnetic particles for at least two reasons. First, it reduces the average density of the suspension and so its sedimentation if, as is generally the case, the density of the non-magnetic particles is much smaller than the one of iron. Second, a low amount of magnetic particles is associated with a lowl magnetic permeability which allows to minimize the demagnetizing field in geometries where the field is perpendicular to a thin sample of the MR suspension. It is also interesting to note that the mechanisms that control viscosity are very different from those operating in conventional MR fluid. In the latter, the magnetic field essentially produces a network of particles whose rigidity is governed by the magnetic force between particles. In this case, it is the yield stress of the suspension which is monitored by the field and there is fairly good agreement between the predictions of the yield stress versus the field and the experimental result. In the case of the bidisperse suspension the yield stress is still given by the magnetic force between particles inside the aggregates of iron particles whatever the size of these aggregates; this is the reason why it remains small and not very different between $F_{CI}$=0.05 and $F_{CI}$=0.29 (cf Fig;(8). In the case of the DST transition with the field, the situation is completely different since the magnetic force only triggers the transition but does not control the rigidity of the resulting aggregated network of particles. One could believe, as it was the case with the jamming theory developed in (Bossis et al., 2022) and following papers, that the suspension could be completely jammed and stop to flow at least until a given yield stress able to break the network of frictional contacts between particles is applied. Experimentally, the suspension continues to flow instead of jamming as can be seen for example in Fig.(9) and also in Fig.(4) for bidisperse suspensions. This is a quite general behavior found on different types of suspension which can be explained by the fact that the arrest of the suspension is an unstable state regarding the polymer layer which was coating the particles before being ejected by the shearing flow. When the flow stops, the hydrodynamic force which was able to eject this stabilizing layer is replaced by a frictional force, but at the molecular-or ionic- level the thermodynamic forces are always present which will bring back the coating layer on the surface of the particles in the absence of flow, thus destroying the frictional contacts and finally allowing to restart the flow. This mechanism well explains the existence of an average constant shear rate above the transition (Bossis et al., 2022). In this case we have an apparent viscosity which grows proportionally to the applied stress above the transition. Below the transition, the effect of the field is mainly to increase the viscosity as can be seen in Fig. (2) for the monodisperse case and in Fig. (4) and table I for the bidisperse case. This very strong increase of the viscosity with the magnetic field is not observed in classical MR fluid. Here we assume that it is due to the interpenetration of the layers of polymer which, at high volume fraction, are already everywhere touching each other during the shear and are still more constrained by the attractive magnetic forces between the particles. So far, we have not measured the sedimentation rate of these suspensions, but we have found that they redisperse within minutes on a vortex shaker after being left to stand for more than 2 years in a bottle. The critical stress and shear rate were respectively 28% and 10% lower than the new sample but still within the uncertainty range of these quantities. For these suspensions to be of interest for applications in shock absorbers where shear rates and stresses are high, it would be necessary to ensure that the particles do not crumble under these conditions. At this stage, it can only be said that, after several stress ramps going well beyond the critical stress, we have not seen any evolution of the rheology. Also we think that, even if under high stresses the largest CC particles could rupture, it would not change their ability to transmit the stress and produce a network of frictional contacts since this phenomenon is not sensitive to the size and shape of particles in the micron size range (Egres and Wagner, 2005). This is nevertheless a point that deserves to be confirmed by high stress experiments over a long time. In practice, because of the complex behavior of the coating layer in the presence of a shear flow and of a magnetic field at these high volume fractions, the monitoring of the viscosity of such suspension can be more complex than with an usual MR fluid but the magnitude of the viscosity change and the low field needed to trigger the transition make these new fluids very attractive for robotic and damping applications.

## ACKNOWLEDGMENT

The authors want to thank the CENTRE NATIONAL D'ETUDES SPATIALES (CNES, the French Space Agency) for having supported this research.

## REFERENCES

Ashtiani, M., Hashemabadi, S.H., Ghaffari, A., 2015. A review on the magnetorheological fluid preparation and stabilization. Journal of magnetism and Magnetic Materials 374, 716–730.

Bossis, G., Boustingorry, P., Grasselli, Y., Meunier, A., Morini, R., Zubarev, A., Volkova, O., 2017. Discontinuous shear thickening in the presence of polymers adsorbed on the surface of calcium carbonate particles. Rheologica Acta 56, 415–430.

Bossis, G., Ciffreo, A., Grasselli, Y., Volkova, O., 2023. Analysis of the rheology of magnetic bidisperse suspensions in the regime of discontinuous shear thickening. Rheol Acta 62, 205–223. https://doi.org/10.1007/s00397-023-01388-x

Bossis, G., Grasselli, Y., Meunier, A., Volkova, O., 2016. Outstanding magnetorheological effect based on discontinuous shear thickening in the presence of a superplastifier molecule. Applied Physics Letters 109, 111902.

Bossis, G., Grasselli, Y., Volkova, O., 2022. Discontinuous Shear Thickening (DST) transition with spherical iron particles coated by adsorbed brush polymer. Physics of Fluids 34, 113317.

Bossis, G., Volkova, O., Grasselli, Y., Ciffreo, A., 2019. The Role of Volume Fraction and Additives on the Rheology of Suspensions of Micron Sized Iron Particles. Front. Mater. https://doi.org/10.3389/fmats.2019.00004

Burguera, E.F., Love, B.J., Sahul, R., Ngatu, G., Wereley, N.M., 2008. A physical basis for stability in bimodal dispersions including micrometer-sized particles and nanoparticles using both linear and non-linear models to describe yield. Journal of intelligent material systems and structures 19, 1361–1367.

Chaudhuri, A., Wang, G., Wereley, N.M., Tasovksi, V., Radhakrishnan, R., 2005. Substitution of micron by nanometer scale powders in magnetorheological fluids, in: Electrorheological Fluids And Magnetorheological Suspensions (Ermr 2004). World Scientific, pp. 391–397.

Chauhan, V., Kumar, A., Sham, R., 2024. Magnetorheological fluids: a comprehensive review. Manufacturing Review 11, 6.

Choi, C., Thompson, R., Hwang, B., 2024. A review of the fabrication of enhanced-performance magnetorheological fluids using carbonyl iron powder. Journal of Molecular Liquids 408, 125308.

Choi, S.-B., 2022. Sedimentation Stability of Magnetorheological Fluids: The State of the Art and Challenging Issues. Micromachines 13, 1904. https://doi.org/10.3390/mi13111904

de Vicente, J., Bossis, G., Lacis, S., Guyot, M., 2002. Permeability measurements in cobalt ferrite and carbonyl iron powders and suspensions. Journal of Magnetism and Magnetic Materials 251, 100–108. https://doi.org/10.1016/S0304-8853(02)00484-5

Egres, R.G., Wagner, N.J., 2005. The rheology and microstructure of acicular precipitated calcium carbonate colloidal suspensions through the shear thickening transition. Journal of rheology 49, 719–746.

Freundlich, H., Röder, H.L., 1938. Dilatancy and its relation to thixotropy. Transactions of the Faraday Society 34, 308–316.

Ginder, J.M., Davis, L.C., Elie, L.D., 1996. Rheology of magnetorheological fluids: models and measurements. International journal of modern physics b 10, 3293–3303.

Guo, F., Xu, Z., Gu, J., 2023. Effects of nano-fumed silica and carbonyl iron powder of different particle sizes on the rheological properties of shear thickening fluids. Colloid Polym Sci 301, 539–555. https://doi.org/10.1007/s00396-023-05087-0

Li, W., Nakano, M., Tian, T., Totsuka, A., Sato, C., 2014. Viscoelastic properties of MR shear thickening fluids. Journal of Fluid Science and Technology 9, JFST0019–JFST0019.

López-López, M.T., Kuzhir, P., Lacis, S., Bossis, G., González-Caballero, F., Durán, J.D., 2006. Magnetorheology for suspensions of solid particles dispersed in ferrofluids. Journal of Physics: Condensed Matter 18, S2803.

Mari, R., Seto, R., Morris, J.F., Denn, M.M., 2014. Shear thickening, frictionless and frictional rheologies in non-Brownian suspensions. Journal of Rheology 58, 1693–1724. https://doi.org/10.1122/1.4890747

Morillas, J.R., de Vicente, J., 2020. Magnetorheology: a review. Soft Matter 16, 9614–9642.

Nejatpour, M., Saleh, M.K.A., Ulasyar, A., Unal, U., Lazoğlu, İ., Acar, H.Y., 2024. Bidisperse magnetorheological fluids with strong magnetorheological response, long-term stability and excellent in-use performance. Smart Materials and Structures 33, 035023.

Pednekar, S., Chun, J., Morris, J.F., 2018. Bidisperse and polydisperse suspension rheology at large solid fraction. Journal of Rheology 62, 513–526.

Qian, C., Tian, Y., Fan, Z., Sun, Z., Ma, Z., 2022. Investigation on rheological characteristics of magnetorheological shear thickening fluids mixed with micro CBN abrasive particles. Smart Materials and Structures 31, 095004.

Seto, R., Mari, R., Morris, J.F., Denn, M.M., 2013. Discontinuous shear thickening of frictional hard-sphere suspensions. Physical review letters 111, 218301.

Sokolovski, V., Tian, T., Ding, J., Li, W., 2020. Fabrication and characterisation of magnetorheological shear thickening fluids. Frontiers in Materials 7, 595100.

Tian, Y., Jiang, J., Meng, Y., Wen, S., 2010. A shear thickening phenomenon in magnetic field controlled-dipolar suspensions. Applied Physics Letters 97, 151904.

Ulicny, J.C., Snavely, K.S., Golden, M.A., Klingenberg, D.J., 2010. Enhancing magnetorheology with nonmagnetizable particles. Appl. Phys. Lett. 96, 231903. https://doi.org/10.1063/1.3431608

Wereley, N.M., Chaudhuri, A., Yoo, J.-H., John, S., Kotha, S., Suggs, A., Radhakrishnan, R., Love, B.J., Sudarshan, T.S., 2006. Bidisperse magnetorheological fluids using Fe particles at nanometer and micron scale. Journal of Intelligent Material Systems and Structures 17, 393–401.

Wilson, B.T., Klingenberg, D.J., 2017. A jamming-like mechanism of yield-stress increase caused by addition of nonmagnetizable particles to magnetorheological suspensions. Journal of Rheology 61, 601–611.

Wyart, M., Cates, M.E., 2014. Discontinuous shear thickening without inertia in dense non-Brownian suspensions. Physical review letters 112, 098302.

Yan, Y., Wei, M., Huang, Z., 2024. Review on shear thickening fluid and its applications in vibration reduction. Materials Research Express 11, 012001.

Zhang, X., Li, W., Gong, X.L., 2008. Study on magnetorheological shear thickening fluid. Smart materials and structures 17, 015051.

Zhao, Q., Wang, K., Yuan, J., Jiang, H., 2023. Research on rheological properties and phenomenological theory-based constitutive model of magnetorheological shear thickening fluids. Smart Materials and Structures 32, 105037.